\documentclass{article}
\pdfoutput=1
\usepackage{amsmath,graphicx,mlspconf}
\usepackage{subfig}

\title{ON THE REPRESENTATION OF SPEECH AND MUSIC}

\name{David N. Levin}

\address{University of Chicago\\
	Department of Radiology\\
	1310 N. Ritchie Ct., Unit 26 AD\\
        Chicago, Illinois 60610}

\begin{document}
\ninept
\maketitle

\begin{abstract}
In most automatic speech recognition (ASR) systems, the audio signal is processed to produce a time series of sensor measurements. This time series encodes semantic information in a speaker-dependent way. An earlier paper showed how to use the sequence of sensor measurements to derive an ``inner" time series that is unaffected by any previous invertible transformation of the sensor measurements. The current paper considers two or more speakers, who mimic one another in the following sense: when they say the same words, they produce sensor states that are invertibly mapped onto one another. It follows that the inner time series of their utterances must be the same when they say the same words. In other words, the inner time series encodes their speech in a manner that is speaker-independent. Consequently, the ASR training process can be simplified by collecting and labelling the inner time series of the utterances of just one speaker, instead of training on the sensor time series of the utterances of a large variety of speakers. A similar argument suggests that the inner time series of music is instrument-independent. This is demonstrated in experiments on monophonic electronic music.
\end{abstract}

\begin{keywords}
speech, music, representation, voice, speaker,\\
instrument
\end{keywords}

\section{INTRODUCTION}
\label{introduction}

Consider a physical system that is being observed with a set of sensors. The time series of raw sensor measurements describes the evolving system in a sensor-dependent way. Appendix and reference \cite{Levin inner} show how a time series of sensor measurements can be statistically processed to create an ``inner" time series that is unaffected by any previous invertible transformation of the sensor measurements. In particular, consider an evolving physical system with $N$ degrees of freedom ($N \geq 1$), and suppose that it is being observed by $N$ sensors, whose output is denoted by $x(t)$ ($x_{k}(t) \mbox{ for } k = 1, \ldots ,N$).
Now, let $x'(t)$ denote another time series of $N$ quantities that is invertibly related to $x(t)$. For example, $x'(t)$ could be the outputs of another set of sensors that are simultaneously observing the same physical system. Appendix and reference \cite{Levin inner} describe how to process $x(t)$ in order to derive an ''inner" time series, $w(t)$ ($w_{k}(t) \mbox{ for } k = 1, \ldots ,N$). Most importantly, it is shown that the \textit{same} inner time series results if the time series, $x'(t)$, is subjected to the same procedure. Notice that this implies that the inner time series is sensor-independent.

Some comments on these results:
\begin{enumerate}
\item In physical terms, an inner time series does not depend on the nature of the sensors, which the listener chose to monitor the system. Instead, it encodes information that is intrinsic to the evolution of the observed system; i.e., information about the nature of the physical system that is ``out there, broadcasting signals to the observer". 
\item In mathematical terms, an inner time series is a self-referential representation of a measurement time series, created by expressing the sensor measurements with respect to their own statistical properties. No matter what linear or nonlinear transformation is applied to a sequence of measurements, its self-referential form (i.e., its inner time series) is the same. In this sense, an inner time series is roughly analogous to the principal component representation of a data set, which displays the data in the same way, no matter what rotation and/or translation has been applied to them. 
\item Appendix and reference \cite{Levin inner} contain a detailed description of the steps required to derive an inner time series from a time series of sensor measurements. In references \cite{Levin inner} and \cite{Levin LVA-ICA}, the sensor-independence of the inner time series is demonstrated with several analytic and numerical examples.
\end{enumerate}

The inner time series may be used to simplify the automatic recognition of speech and music. In its simplest form, speech recognition is performed by compiling a look-up table, which lists each phoneme in the language's vocabulary, together with the corresponding sensor signals \cite{Huang}.  An unknown signal is then recognized by determining which phoneme is associated with the signal that is ``nearest" to the unknown one. In practice, this rudimentary recognition process must be modified to account for the fact that a given phoneme may be represented by many different sensor measurements.  For example, different acoustic signals may result when the same phoneme is uttered by different speakers having different vocal tracts. This makes it necessary to laboriously collect and label many ``training" signals for each phoneme. These training signals are used to construct a statistical model for each phoneme (e.g., a hidden Markov model, \cite{HTK}). The recognition process then involves a probabilistic computation for determining which phoneme model is the most likely source of the unknown signal.

The current paper suggests that the various speaker-dependent renditions of a given phoneme have the \textit{same} inner time series, and this fact can simplify the process of training an ASR system. To see why this may be the case, consider the sensor states, $x(t)$ and $x'(t)$, which are produced when two speakers are saying the same sequence of phonemes. Now, suppose that these two individuals consistently mimic one another when they are saying the same phonemes. In other words, whenever the first speaker produces the sensor state $x$, the second speaker produces the sensor state $x'$. Likewise, whenever the second speaker produces the sensor state $x'$, the first speaker produces the sensor state $x$. It can be argued that this type of mimicry is likely to occur because it describes the way that inexperienced speakers try to imitate the vocal tract gestures of experienced speakers. In any event, if mimicry is present, it defines an invertible transformation that maps $x(t)$ and $x'(t)$ onto one another. As discussed in the first paragraph, it follows that these two measurement time series must have the same inner time series. In short, if all speakers of a given language mimic one another, all of their different renditions of a given phoneme have the same inner time series.

If this type of mimicry occurs, ASR is best performed by representing phonemes by their inner time series, instead of their ``outer" time series (the time series of sensor measurements). Because each phoneme is associated with a single inner time series, each unknown signal can be recognized by simply identifying the phoneme whose inner time series is ``closest" to the inner time series of that unknown signal. There is no need to laboriously collect and label the numerous speaker-dependent ``outer" time series corresponding to each phoneme.

The above-described methodology can also be applied to the representation of music. Specifically, if musicians consistently mimic one another when they play the same notes on different instruments, their audio signals will have the same inner time series. In other words, the inner time series will represent the music in an instrument-independent manner. Monophonic music is the simplest application domain for testing these ideas because its dimensionality is low ($d = 2$), compared to the dimensionality of speech ($d \geq  5$) or polyphonic music ($d \geq 4$). Therefore, in the next section, we demonstrate these ideas by applying them to monophonic music. Specifically, we show how the inner time series of monophonic music encodes it in a way that is both sensor-independent and instrument-independent (the musical analogue of speaker-independent).

\section{EXPERIMENTS WITH MONOPHONIC MUSIC}
\label{experiments}

In the following subsection, approximately 25 minutes of monophonic piano music were processed by two different filterbanks. Then, the two time series of sensor measurements were used to derive a pair of inner time series of the music. The latter were shown to be approximately the same, thereby demonstrating the sensor-independence of the inner time series, as asserted in the first paragraph of this paper. 

In the second subsection, an electronic piano was used to play music in two different keys, effectively simulating music played on two different instruments that are mimicking one another. Those two renditions of the music were used to derive a pair of inner time series, which were approximately the same. Each of these two inner time series also approximated the inner time series derived from a violin rendition of the same music. These results demonstrated the instrument-independence of the inner time series of monophonic music.

\subsection{Sensor-independence of the inner time series}
\label{sensor-independence}

\begin{figure*}
\centering
\subfloat[]{\includegraphics[width=4.0cm]{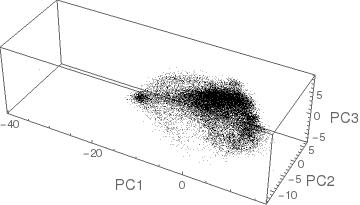}}
\subfloat[]{\includegraphics[width=4.0cm]{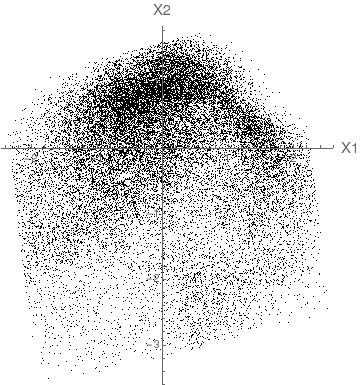}}
\hspace{0mm}
\subfloat[]{\includegraphics[width=4.0cm]{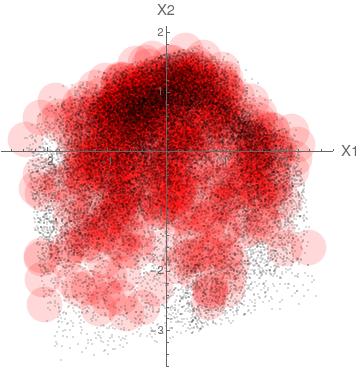}}
\subfloat[]{\includegraphics[width=4.0cm]{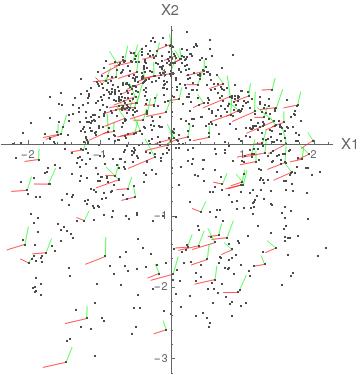}}
\caption{(a) The principal components of typical output of the first filterbank during 25 minutes of monophonic piano music. (b) The distribution of the data in panel a, after they were dimensionally reduced and subjected to variance normalization. (c) The neighborhoods in which the local vectors ($V_{(i)}$) were computed. (d) The red and green lines depict the local vectors ($V_{(1)}$ and $V_{(2)}$, respectively) in the neighborhoods in panel c.}
\label{fig-2x2-8000}
\end{figure*}

\begin{figure*}
\centering
\subfloat[]{\includegraphics[width=4.0cm]{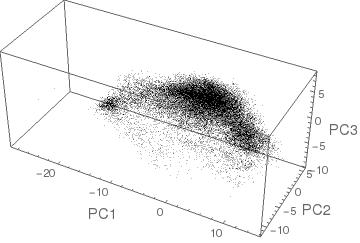}}
\subfloat[]{\includegraphics[width=4.0cm]{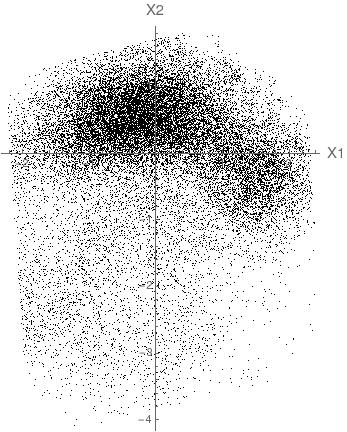}}
\hspace{0mm}
\subfloat[]{\includegraphics[width=4.0cm]{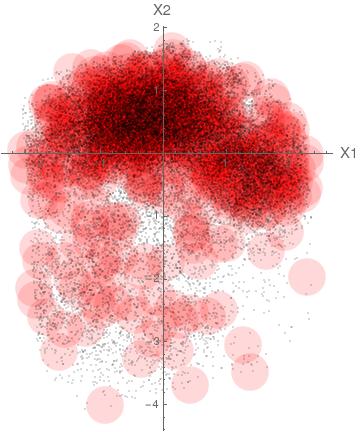}}
\subfloat[]{\includegraphics[width=4.0cm]{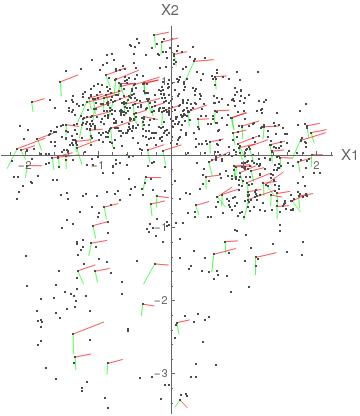}}
\caption{(a) The principal components of typical output of the second filterbank during 25 minutes of monophonic piano music. (b) The distribution of the data in panel a, after they were dimensionally reduced and subjected to variance normalization. (c) The neighborhoods in which the local vectors ($V_{(i)}$) were computed. (d) The red and green lines are the local vectors ($V_{(1)}$ and $V_{(2)}$, respectively) in the neighborhoods in panel c.}
\label{fig-2x2-4000}
\end{figure*}

Approximately 25 minutes of monophonic piano music was created on an electronic keyboard (a Yamaha DGX640 keyboard with a grand piano ``voice"). The music consisted of multiple randomly-chosen ``runs", with each run comprised of a sequence of several dozen notes that were separated by semitone pitch intervals and by approximately 0.4 s time intervals. The resulting audio signal was sampled at 16 kHz with two bytes of depth. These samples were processed with a series of Hamming windows, having 25 ms width and separated by 5 ms time intervals. The data in each window were then Fourier transformed to compute the log power output of 15 Mel-scale filterbank channels (\cite{Huang}, \cite{HTK}), spanning the frequency range: $0 - 8000 Hz$. Figure \ref{fig-2x2-8000}a shows the first three principal components of these multiplets. Notice the tail of points extending to the left. This corresponds to noise during the short periods of silence between runs of notes. This tail was truncated by excluding all points having a low first principal component; namely, $PC1 < -10$. Note that monophonic music is expected to have two degrees of freedom, corresponding to the identity of each played note and its degree of excitation. Therefore, the data were dimensionally reduced by retaining just the first two principal components. Next, outliers were removed by excluding the lowest (and highest) $0.5\%$ of points along each measurement axis. Finally, the remaining distribution was transformed linearly along each axis in order to normalize its variance along each direction. Figure \ref{fig-2x2-8000}b shows the resulting distribution of multiplets, which were taken to be the sensor measurements, $x(t)$. Figure \ref{fig-2x2-8000}c shows a sample of approximately 1000 disk-shaped local neighborhoods that covered the distribution of sensor measurements. In each of these neighborhoods, the local vectors ($V_{(i)}$ for $i = 1,2$) were calculated, as described in Appendix and in reference \cite{Levin inner}. Figure \ref{fig-2x2-8000}d shows the local vectors computed in these neighborhoods, after rescaling them by a factor of two for the purpose of display. These vectors were used to compute the inner time series ($w_{i}(t)$ for $i = 1,2$) of the sensor measurements in a typical 1.5 s interval. This was done by substituting the local vectors ($V_{(i)}$) and the local measurements ($x(t)$) into Eq.(8) of Appendix. The solid black line in figure \ref{fig-FBComp} shows the inner time series after high frequency noise was removed by Gaussian filtration with a standard deviation equal to 2.5 frames.

A second set of measurements was created by following the above procedure, except that the Mel-scale filterbank had only 13 channels, which covered a reduced frequency range ($300 - 4000 Hz$). Figures \ref{fig-2x2-4000}a-d show the principal components of the filterbank outputs, the distribution of sensor measurements, the neighborhoods in which the local vectors were computed, and the local vectors, respectively. The dashed red line in Figure \ref{fig-FBComp} shows the inner time series that was calculated on the same 1.5 s time interval depicted by the solid black line. As discussed in Appendix and in reference \cite{Levin inner}, the two inner time series, which were derived from the outputs of different filterbanks, were expected to be the same, except for a global reflection or permutation transformation ($P$). Figure \ref{fig-FBComp} compares the two inner time series after the second series was transformed by a global reflection in the origin. After that adjustment, it was apparent that the two inner time series were quite similar, even though they were derived from measurements made with different sensors (i.e., different filterbanks). This demonstrates that the inner time series is sensor-independent. It also shows that the inner time series is listener-independent, in the sense that it does not depend on the listener's choice of sensors for observing the system.

\begin{figure}
\centering
\subfloat{\includegraphics[width=1.0\linewidth]{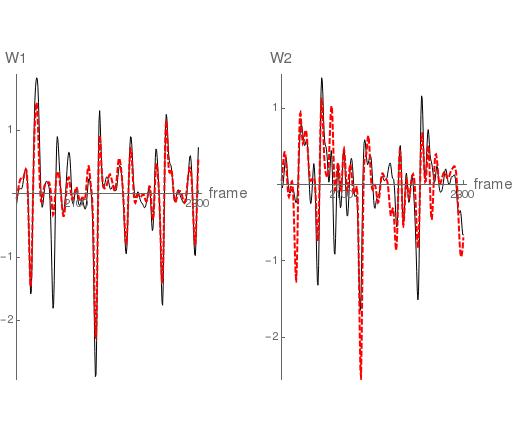}}
\caption{The solid black and dashed red lines show the inner time series of music during a 1.5 s time interval, computed from the filterbank measurements in Figures \ref{fig-2x2-8000}b and  \ref{fig-2x2-4000}b, respectively.}
\label{fig-FBComp}
\end{figure}

\subsection{Instrument-independence of the inner time series}
\label{Instrument-independence}

\begin{figure}
\centering
\subfloat{\includegraphics[width=1.0\linewidth]{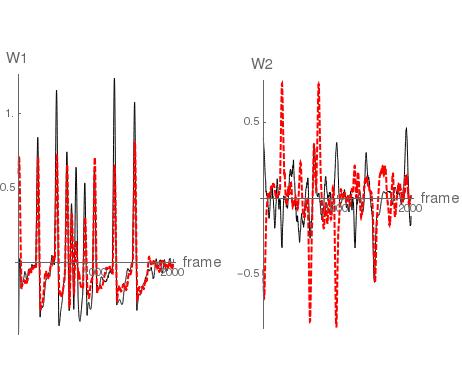}}
\caption{The solid black and dashed red lines show the inner time series of 10 s of monophonic music (the first few measures of ``Over the Rainbow"), played without and with upward transposition by 6 semitones, respectively.}
\label{fig-piano-transPiano}
\end{figure}

In another experiment, the first of the above-described filterbanks was used to compute a time series of sensor measurements from monophonic piano music, consisting of 12.5 minutes of semitone ``runs", as well as 10 s of the tune,``Over the Rainbow". The inner time series of this music was computed by substituting those sensor measurements and the corresponding local vectors into Eq.(8) of Appendix. The result is shown by the solid black line in Figure \ref{fig-piano-transPiano}, after Gaussian smoothing with standard deviation equal to 12.5 frames. Next, the same music was electronically transposed to a key that was six semitones higher. This rendition was also used to compute a time series of sensor measurements from the first filterbank's outputs. These measurements and the corresponding local vectors were substituted into Eq.(8) of Appendix in order to compute the inner time series of the transposed music. The result is shown by the dashed red line in Figure \ref{fig-piano-transPiano}. The similarity of the two inner time series in this figure demonstrates that the inner time series is approximately transposition-independent. Note that a musical instrument, which consistently plays transposed scores, can be considered to be another instrument, which is different from the one that plays untransposed scores. In this sense, Figure \ref{fig-piano-transPiano} demonstrates that the inner time series is independent of the utilized musical instrument.

\begin{figure}
\centering
\subfloat{\includegraphics[width=1.0\linewidth]{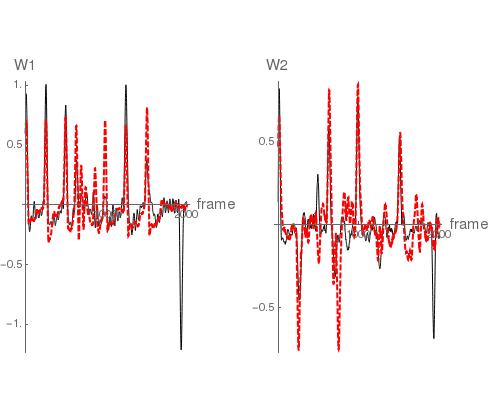}}
\caption{The solid black and dashed red lines show the inner time series of 10 s of monophonic music (the first few measures of ``Over the Rainbow"), played with the voice of an electronic violin and a ``transposed electronic piano", respectively.}
\label{fig-violin-transPiano}
\end{figure}

\begin{figure}
\centering
\subfloat{\includegraphics[width=1.0\linewidth]{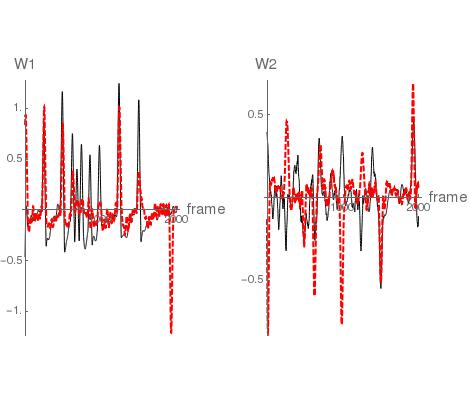}}
\caption{The solid black and dashed red lines show the inner time series of 10 s of monophonic music (the first few measures of ``Over the Rainbow"), played with the voices of an electronic piano and an electronic violin, respectively.}
\label{fig-piano-violin}
\end{figure}

In another experiment, the procedure in the last paragraph was applied to the same monophonic score, after it was electronically transformed to have a violin-like voice. As described above, the audio signals for this music were used to compute: 1) the time series of sensor measurements, derived from the first filterbank's outputs; 2) the corresponding inner time series, shown as the solid black line in Figure \ref{fig-violin-transPiano}. The dashed red line in Figure \ref{fig-violin-transPiano} shows the inner time series derived from the same music, rendered in the voice of the above-described ``transposed piano". It is evident that the two inner time series are approximately the same. This demonstrates again that an inner time series represents monophonic music in a largely instrument-independent manner. Instrument-independence of the inner time series is also suggested by Figure \ref{fig-piano-violin}, which compares the inner time series of an electronic piano and an electronic violin, respectively, when they play the same music.

Some comments on these results:
\begin{enumerate}
\item As noted in Appendix and in reference \cite{Levin inner}, the local vectors and the inner time series are only determined up to a global permutation or reflection transformation. Therefore, this type of transformation was applied to the inner time series in Figures \ref{fig-piano-transPiano} - \ref{fig-piano-violin} in order to maximize the agreement of the two inner time series depicted in each figure.
\item Without additional experiments, it is difficult to determine the physical meaning of the two components of the inner time series. However, there is a suggestive clue: in Figures \ref{fig-piano-transPiano} - \ref{fig-piano-violin}, the $w_{1}$ components are noticeably more rendition-independent (more instrument-independent) than are the $w_{2}$ components. Recall that the instrument-independence of an inner time series reflects how accurately each instrument mimicked the other one. This suggests that $w_{1}$ encodes information that is easy to mimic accurately from one rendition to the other; for example, it may encode frequency information, which simply reflects what notes were played. Likewise, it suggests that $w_{2}$ encodes information that is more difficult to mimic accurately from one rendition to the other. For example, it may encode loudness information that reflects the degree of excitation of each note, something which is more difficult to control, especially on an electronic keyboard.
\end{enumerate}

\section{CONCLUSION}
\label{conclusion}

Most methods of automatic speech recognition represent speech as a time series of audio sensor measurements, such as filterbank outputs. These measurements depend on: 1) the semantic content of the speech signal; 2) the type of sensors that are utilized by the listener (e.g., the type of filterbank); 3) the speaker's vocal characteristics (e.g., the nature of his/her vocal tract). Because of the speaker-dependence of the measurement time series, a typical ASR system must be trained on a representative sample of the many signals that can represent each phoneme. In general, the collection and labelling of these training signals is the most laborious step in creating ASR software.

Appendix and \cite{Levin inner} showed how to use the statistical properties of a measurement time series to derive an ``inner" time series that is unaffected by any prior invertible, possibly nonlinear, point transformation of the measurements. This implies that the inner time series is independent of the type of sensors chosen by the listener.

In the current paper, we assume that all pairs of speakers of a given language (e.g., speakers $S$ and $S'$) mimic each other when they are saying the same thing. This assumption is reasonable, given the fact that people seem to learn to speak by imitating experienced speakers. Mathematically, this means that there is an invertible transformation that relates the sensor states ($x(t)$) produced by speaker $S$ and the sensor states ($x'(t)$) produced by any other speaker $S'$, who is saying the same words. It follows from the results in Appendix and \cite{Levin inner} that the two measurement time series ($x(t)$ and $x'(t)$) have the same inner time series. This means that the inner time series is speaker-independent, as well as listener-independent. Evidently, the inner time series is only dependent on the semantic content of an utterance. 

If the assumption of mimicry is valid, the ASR process can be significantly simplified by representing speech with its inner time series, instead of its measurement time series. Specifically, instead of representing each phoneme by a distribution of speaker-dependent measurement time series, each phoneme can be represented by a single speaker-independent inner time series. Thus, it is not necessary to laboriously collect and label the many ``outer" time series corresponding to each phoneme. Instead, it suffices to know the unique inner time series that represents each phoneme. At least in principle, phoneme recognition can then be performed by identifying the phoneme having the inner time series that is ``nearest" to the inner time series of the unknown signal.

Analogous ideas are expected to apply to music. In particular, the inner time series of music is expected to be instrument-independent, as long as musicians mimic one another when they are playing the same notes. And, in fact, the experiments in Section 2 show that monophonic music appears to have this important property. Of course, the next step is to perform similar experiments on polyphonic music and speech.

\section{Appendix}
\label{appendix}

This appendix describes how a time series of sensor measurements ($x(t)$) can be processed to derive a sensor-independent representation of the data, called the ``inner" time series. An outline of this procedure is presented here; a detailed description can be found in \cite{Levin inner} and \cite{Levin LVA-ICA}.

 The first step is to construct second-order and fourth-order local correlations of the data's velocity ($\dot{x}$)
\begin{equation}
\label{C2 definition}
C_{kl}(x) = \, \langle (\dot{x}_k-\bar{\dot{x}}_k) (\dot{x}_l-\bar{\dot{x}}_l) \rangle_{x} 
\end{equation}
\begin{equation}
\label{C4 definition}
C_{klmn}(x) = \, \langle (\dot{x}_k-\bar{\dot{x}}_k) (\dot{x}_l-\bar{\dot{x}}_l) 
(\dot{x}_m-\bar{\dot{x}}_m) (\dot{x}_n-\bar{\dot{x}}_n) \rangle_{x}
\end{equation}
where $\bar{\dot{x}} = \langle \dot{x} \rangle_x$, where the bracket denotes the time average over the trajectory's segments in a small neighborhood of $x$, and where all subscripts are integers between $1$ and $N$ with $N \geq 1$.

Next, let $M(x)$ be any local $N \times N$ matrix, and use it to define $M \mbox{-transformed}$ velocity correlations, $I_{kl}$ and $I_{klmn}$
\begin{equation}
\label{I2 definition}
I_{kl}(x) = \sum_{1 \leq k', \, l' \leq N} M_{kk'}(x) M_{ll'}(x) C_{k'l'}(x) ,
\end{equation}
\begin{equation}
\label{I4 definition}
\begin{split}
I_{klmn}(x) = \sum_{1 \leq k', \, l', \, m', \, n' \leq N} M_{kk'}(x) M_{ll'}(x) \\
M_{mm'}(x) M_{nn'}(x) C_{k'l'm'n'}(x) .
\end{split}
\end{equation}
Because $C_{kl}(x)$ is generically positive definite at any point $x$, it is almost always possible to find a form of $M(x)$ that satisfies
\begin{equation}
\label{M definition 1}
I_{kl}(x) = \delta_{kl}
\end{equation}
\begin{equation}
\label{M definition 2}
\sum_{1 \leq m \leq N} I_{klmm}(x) = D_{kl}(x) , 
\end{equation}
where $D(x)$ is a diagonal $N \times N$ matrix. If $D$ is not degenerate, $M(x)$ is unique, up to arbitrary \textit{local} permutations and/or reflections. In almost all applications of interest, the velocity correlations will be continuous functions of $x$. Therefore, in any neighborhood, there will always be a continuous solution for $M(x)$, and this solution is unique, up to arbitrary \textit{global} permutations and/or reflections. 

In any other coordinate system $x'$, the most general solution for $M'$ is given by
\begin{equation}
\label{M'}
M'_{kl}(x') = \sum_{1 \leq m, \, n \leq N} P_{km} M_{mn}(x) \frac{\partial x_n}{ \partial x'_l} ,
\end{equation}
where $M$ is a matrix that satisfies (\ref{M definition 1}) and (\ref{M definition 2}) in the $x$ coordinate system and where $P$ is a product of permutation, reflection, and identity matrices (\cite{Levin LVA-ICA}).

Notice that (\ref{M'}) shows that the rows of $M$ transform as local covariant vectors, up to a global permutation and/or reflection. Likewise, the same equation implies that the columns of $M^{-1}$ transform as local contravariant vectors (denoted as $V_{(i)}(x) \mbox{ for } i = 1, \ldots N$), up to a global permutation and/or reflection. Because these vectors are linearly independent, the measurement velocity at each time ($\dot{x}(t)$) can be represented by a weighted superposition of them
\begin{equation}
\label{xDot rep}
\dot{x}(t) = \sum_{1 \leq i \leq N} w_{i}(t) V_{(i)}(x)  ,
\end{equation}
where $w_{i}$ are time-dependent weights. Because $\dot{x}$ and $V_{(i)}$ transform as contravariant vectors (except for a possible global permutation and/or reflection), the weights $w_{i}$ must transform as scalars or invariants; i.e., they are independent of the coordinate system in which they are computed (except for a possible permutation and/or reflection). Therefore, the time-dependent weights, $w_{i}(t)$, provide an inner (coordinate-system-independent) description of the system's velocity in measurement space. Two observers, who use different sensors will derive the same inner time series, except for a possible global permutation and/or reflection.

A technical comment: In order to compute the inner time series ($w_{i}(t)$) along a path $x(t)$, it is necessary to compute the local vectors there ($V_{(i)}[x(t)]$). To do this, the measurement time series must densely sample the neighborhoods near that path so that the local fourth- and lower-order correlations of $\dot{x}$ can be computed there. Higher-order statistical quantities need not be computed.

\vfill
\pagebreak

\bibliographystyle{IEEEbib}
\bibliography{strings,refs}

\begin{thebibliography}{1} 
\bibitem{Levin inner} D.~N. Levin, ''The Inner Structure of Time-Dependent Signals," http://arxiv.org/abs/1703.08596 (March 24, 2017).

\bibitem{Levin LVA-ICA} D.~N. Levin, ''Model-independent method of nonlinear blind source separation," In: \textit{Tichavsky, P., Babaie-Zadeh, M., Michel, O., and Thirion-Moreau, N. (eds.), Latent Variable Analysis and Signal Separation, Lecture Notes in Computer Science, Springer}, vol. 10169, pp. 310-319, 2017.

\bibitem{Huang} X.~Huang, A.~Acero, and H-W.~Hon, \textit{Spoken Language Processing.} Upper Saddle River, New Jersey: Prentice Hall PTR, 2001.

\bibitem{HTK} S.~Young et al, \textit{The HTK Book.} Cambridge University Engineering Department, Cambridge, England, 2005. 
\end{thebibliography}

\end{document}